\documentclass[12pt]{article}

\usepackage[ansinew]{inputenc}
\usepackage[english]{babel}
\usepackage{amsmath,amsfonts,amssymb,mathrsfs,cancel}
\usepackage[dvips]{graphicx}

\usepackage{epsfig}
\usepackage{graphicx}
\usepackage{subfigure}
\usepackage{cite}

\begin{document}

\newcommand{\be}{\begin{equation}}\newcommand{\ee}{\end{equation}}
\newcommand{\bea}{\begin{eqnarray}} \newcommand{\eea}{\end{eqnarray}}
\newcommand{\ba}[1]{\begin{array}{#1}} \newcommand{\ea}{\end{array}}

\long\def\symbolfootnote[#1]#2{\begingroup%
\def\thefootnote{\fnsymbol{footnote}}\footnote[#1]{#2}\endgroup} 

\numberwithin{equation}{section}

\def\pd{\partial}
\def\p{\partial}
\def\a{\alpha}
\def\b{\beta}
\def\m{\mu}
\def\mm{m}
\def\n{\nu}
\def\r{\rho}
\def\s{\sigma}
\def\d{\delta}
\def\wg{\wedge}
\def\eps{\epsilon}
\def\veps{\varepsilon}
\def\nn{\nonumber}
\def\ov{\over }
\def\td{\tilde }
\def\Omm{\mu} 

\rightline{ICCUB-12-131}

\rightline{April, 2012}

\bigskip

\begin{center}

{\Large\bf  
Supersymmetric  BCS 
}

\bigskip
\bigskip

{\it \large Alejandro Barranco \footnotemark[1] and Jorge G. Russo\symbolfootnote[1]{On leave of absence from Universitat de Barcelona and Institute of Cosmos Sciences, Barcelona, Spain.}\footnotemark[2]\footnotemark[3]}
\bigskip

\end{center}

{\it

\begin{enumerate}

\item  Institute of Cosmos Sciences and ECM Department, Facultat de F{\'\i}sica\\
 Universitat de Barcelona, Av. Diagonal 647,  08028 Barcelona, Spain

\item  Perimeter Institute for Theoretical Physics,\\
Waterloo, Ontario, N2L 2Y5, Canada

\item Instituci\'o Catalana de Recerca i Estudis Avan\c cats (ICREA), \\
Pg. Lluis Companys, 23, 08010 Barcelona, Spain

\end{enumerate}

}
\bigskip
\bigskip

\begin{abstract}

We implement relativistic BCS superconductivity in $\mathcal N=1$ supersymmetric field theories with a $U(1)_R$ symmetry. 
The simplest model contains two chiral superfields with  a  K\"ahler potential modified by quartic terms.
We study the  phase diagram  of the gap as a function of the temperature and the specific heat.
The superconducting phase transition turns out to be first order,
 due to the scalar contribution to the one-loop potential.
By virtue of supersymmetry, the critical curves depend logarithmically with the UV cutoff,
rather than quadratically as in standard BCS theory. 
We comment on the difficulties in having fermion condensates when the chemical potential
is instead coupled to a baryonic $U(1)_B$ current.
We also discuss supersymmetric models of BCS
with canonical K\"ahler potential  constructed by 
``integrating-in" chiral superfields.


\end{abstract}

\clearpage

\tableofcontents

\section{Introduction}

 Superconductivity is a common phenomenon that arises whenever there is spontaneous symmetry breaking (SSB) of a local
$U(1)$ symmetry. BCS is a particular theory realizing SSB, where 
one starts with a theory with a local $U(1)$ symmetry and quantum effects at finite chemical
potential generate a SSB vacuum by fermion condensation. 
The IR choice of vacuum
can be described in terms of an effective Landau-Ginzburg theory (which can
be derived from BCS). The low energy excitation spectrum can be described in terms of a Landau liquid, where the excitations are fermions.

The plan of this work is to investigate the extent to which the dynamics of relativistic BCS theory \cite{Bailin:1983bm,Bertrand}, with its usual features, can be implemented
within the context of ${\cal N}=1$ supersymmetric field theory. 
A supersymmetric model for  chiral symmetry breaking produced by fermion condensation
at zero temperature and zero chemical potential was discussed in \cite{Buchmuller:1982ty}.
This model was generalized in \cite{Ohsaku:2005rb} to incorporate  BCS type superconductivity, but the construction uses explicit supersymmetry breaking terms --therefore the Lagrangian does not describe a supersymmetric theory (in addition, it involves approximations where some terms of the Lagrangian need to be neglected). To our knowledge, there has been no discussion  in the literature implementing BCS superconductivity in supersymmetric theories.

BCS requires the introduction of chemical potential for the fermions and in supersymmetric theories this leads to some obvious problems. Consistency demands that this chemical potential be coupled to 
a (non-anomalous) $U(1)$ current. For a baryonic $U(1)_B$ symmetry, in supersymmetric theories, this can only be done in a consistent way  by simultaneously introducing  the same chemical potential for the scalars.
But charged  scalar fields in the presence of chemical potential
can run into problems of Bose-Einstein (BE) condensation when the chemical potential becomes greater than the mass. 
 The one-loop potential becomes ill-defined due to divergences. 
Adding a mass term to the superpotential does not circumvent this problem
because the requirement of existence of Fermi surfaces, due to the relations between mass parameters implied by supersymmetry, is always correlated to the
appearance of BE condensation.
We will evade this problem by coupling the chemical potential to a $U(1)_R$ current and considering    models where the light scalars have vanishing $U(1)_R$ charge.


Another  approach (used in \cite{Yamada:2006rx,Hollowood:2008gp}) is to compute thermodynamics quantities for the theory on  $S^1\times S^3$,  where the three-sphere has radius $R$ and the scalar fields have a
 mass equal to $1/R$.  Then the free energy of the system can be computed in a certain regime of parameters, typically, for chemical potentials which are lower than (or equal
to) $1/R$. However, we will see that in this approach the scalar mass scale cannot be separated from the Fermi energy.
Although this does not completely preclude the construction of models with  fermion condensates,
it nevertheless implies that any model of this sort will be on the verge of producing BE condensation by a slight modification of parameters. It also implies that the vacuum dynamics will be governed not only by fermions
near the Fermi surface but it will also be strongly affected by the scalar field fluctuations, which in some cases can be
dominant.


Some previous studies of phase transitions in supersymmetric field theories have not found   superconducting phases.
 In particular, in \cite{Hollowood:2008gp}, the free energy for $\mathcal
 N=4$ super Yang-Mills theory
was computed in detail for two particular values of the chemical potential, $\mu_i =0$ or $\mu_i=1/R$, $i=1,2,3$ (associated with $U(1)\times U(1)\times U(1)\subset SO(6)_R$). 
However, so far no sign  of a superconducting phase transition was found \cite{Hollowood:2008gp}, despite all rich phenomena that seem to be taking place on the gravity side at strong coupling
\cite{Aprile:2011uq}. Another detailed search for $U(1)$ breaking transitions was carried out in \cite{Ammon:2011hz} for ${\cal N}=4$ supersymmetric Yang-Mills theory coupled to a
single massive fundamental-representation $\mathcal N = 2$ hypermultiplet, but no evidence of any instability was found.
Since any theory with higher supersymmetry can be viewed as a particular $\mathcal N=1$ supersymmetric field theory,
it is convenient to use the $\mathcal N=1$ framework to provide a general picture of the conditions under which
$U(1)$ SSB can arise by BCS fermion condensation.

This paper is organized as follows. In section 2 we briefly review relativistic BCS theory,
as this will provide the basis for the construction of a BCS theory in the supersymmetric case.
In section 3 we construct  $\mathcal N=1$ supersymmetric Lagrangians with quartic fermion interactions.
In section 3.1 we discuss a model with Fermi surfaces but with problems of BE condensation.
We show that the one-loop potential is ill-defined even in $S^1\times S^3$  as soon as Fermi surfaces appear.
In section 3.2 we present a simple example of an $\mathcal N=1$ supersymmetric theory that exhibits  BCS superconductivity
with no problems of BE condensation arising from the scalar sector.
In section 4 we consider models with canonical K\" ahler potential, and discuss the difficulties in implementing
BCS superconductivity in these type of theories if all fields are dynamical and  interactions exclusively arise from the superpotential.
Section 5 summarizes the results and Appendices A and B contain further details of the calculations.

\section{Relativistic BCS theory}

 It is useful to briefly review the main features of relativistic BCS theory. Here we will work with global $U(1)$ symmetries,
so in this sense we will be discussing superfluidity, although  transport properties are similar in both cases.
In relativistic BCS theory, one has the effective Lagrangian \cite{Bertrand} 
\be
\mathcal L={i\over 2} (\bar   \psi \gamma^\mu \partial_\mu \psi -\partial_\mu \bar   \psi \gamma^\mu \psi)-m\bar   \psi \psi+\mu\psi^\dagger\psi+ {g^2\over 2}(
\bar   \psi_c \gamma_5 \psi )^\dagger (
\bar   \psi_c \gamma_5 \psi )\ .
\ee
The $U(1)$ symmetry ensures fermion number conservation, which allows the introduction of  the chemical potential in the usual way.
The Lagrangian is not renormalizable, the four-fermion interaction typically represents an irrelevant operator, but the dynamics of BCS superconductivity is such that
for fermions which are close to the Fermi surface this attractive, four-fermion interaction becomes strong.
At weak coupling the scaling dimension of the
fermionic fields must be very close to that of
the 3/2 for a free field. Hence, on dimensional
grounds, the interaction term is irrelevant in the IR.
Naively it would seem that this theory cannot lead to
any interesting IR physics. The phenomenon that actually takes place is
 explained in \cite{Polchinski:1992ed}. The key observation
is that, in the presence of a chemical potential
there is a Fermi surface which can change the
naive scaling dimensions for the operators in 
such a way that the
otherwise irrelevant interaction becomes indeed
marginal. This is the seed for the possibility of
a non-trivial IR physics such as superconductivity.

Technically, to study this system, one considers the Euclidean theory at finite temperature and performs 
 a Hubbard-Stratonovich transformation, one introduces the auxiliary field $\Delta(x)$ and the Lagrangian is then replaced by
\bea
\mathcal L_E &=&{1\over 2} (\psi^\dagger  \partial_\tau \psi -\partial_\tau \psi^\dagger \psi)
-{i\over 2} (\bar   \psi \gamma^i \partial_i \psi -\partial_i \bar   \psi \gamma^i\psi )
+m\bar   \psi \psi-\mu\psi^\dagger\psi
\nn\\
&+& {1\over 2g^2} |\Delta|^2 -{1\over 2} \big[ \Delta^\dagger  (
\bar   \psi_c \gamma_5 \psi )+     \Delta (\bar   \psi_c \gamma_5 \psi )^\dagger \big]\ .
\label{bcsl}
\eea
The Lagrangian becomes quadratic in the fermions, which can now be integrated out explicitly
giving rise to an effective potential for $\Delta $.
The fermion energy eigenvalues are 
\be
\omega_\pm=\sqrt{(\omega_0(\vec p) \pm \mu)^ 2+|\Delta|^2}\ ,\qquad \omega_0\equiv \sqrt{p^2+m^2}\ ,
\ee
where $\pm $ stands for particles and antiparticles. The one-loop effective potential $\Omega $
is then obtained as usual by adding to the classical potential, ${1\over 2g} |\Delta|^2$,  the thermal contribution
  $$ 
-{2\over\beta} \int \frac{d^ 3p}{(2\pi)^ 3} \left(  \log(1+e^{-\beta \omega_-(p)})+ \log(1+e^{-\beta \omega_+(p)})\right) \ ,
$$ 
plus a  (Coleman-Weinberg) contribution that survives at zero temperature,
\be
\int \frac{d^ 3p}{(2\pi)^ 3} \left(2\omega_0(p)-  \omega_-(p)- \omega_+(p)\right)\ .
\ee
$\Omega $ is the thermodynamic potential of the grand canonical ensemble.
The integral over momentum  for this contribution is divergent. Since the theory is not renormalizable,  one must restrict  to energies below a cutoff $\Lambda$ (``Debye" energy).
The cutoff as usual represents the energy where new physics emerges.
At low temperatures, the dominant contribution then arises from frequencies $\omega_0 $ near $\mu $ and the  contribution of the antiparticle --represented by the terms with $\omega_+$-- can be neglected (we assume $\mu>0$; if $\mu<0$ it is the antiparticle contribution the dominant one).
Since $p^2>0$, the existence of a Fermi surface at a finite momentum $p_F$
 requires $\mu>m$, where $p_F$ is defined by the condition $\sqrt{p_F^2+m^2}=\mu$.
As a result the system has a Fermi energy represented by the  chemical potential $\mu$.
If $\Delta =0$, this represents the Fermi energy in the usual sense,
at zero temperature fermions would occupy energy levels with
$\omega_0(p)<\mu $. 
However, in this system, $\Delta $ is spontaneously turned on  below some critical temperature.
When $\Delta $ is not equal to zero, there is a fermion condensate and  
the energy eigenvalues  $\omega_\pm (p)$ do not vanish at any value
of momentum.  At low temperatures, 
the dominant contributions
come from the region where $\omega_-(p) $ has a minimum value. The location in momentum space of this  minimum value
 defines the concept of Fermi surface in more general situations. For this system,
this still occurs at $\omega_0(p)=\mu$, though we will see that in more general systems the Fermi surface location can be shifted when $\Delta $ is turned on. 

The instability leading to $\Delta\neq 0$ and thus to the formation of the fermion condensate appears when the coefficient of the $O(\Delta^2 )$ term in the complete expression for the one-loop effective potential changes sign. Let us examine the conditions under which the fermion condensate appears.
Expanding the full one-loop effective potential $\Omega$ including the Coleman-Weinberg and thermal part
in powers of $\Delta $, one finds
\be
\Omega\big|_{\Delta^2}={\Delta^2\over 2g^2}   \bigg( 1- {g^2\over 2\pi^2}\int_{0}^\Lambda dp\ p^2 \Big( \frac{\tanh( \frac{1}{2} \beta (\omega_0(p)-\mu))}{ \omega_0(p)-\mu} + \frac{\tanh( \frac{1}{2} \beta (\omega_0(p)+\mu))}{ \omega_0(p)+\mu}\Big)\bigg)\ .
\label{doscinco}
\ee
 The equation $\Omega\big|_{\Delta^2}=0$  determines the critical temperature for the formation of the fermion condensate. As long as $\mu>m$ this equation always defines a finite critical temperature for any value
of $g,\ m,\ \mu$ and $\Lambda\gg m$.
The gap equation $\Delta=\Delta(T )$ is  obtained by
differentiating the one-loop effective potential with respect to $\Delta $.
We find
\be
1 =\frac{g^2 }{2\pi ^2}\int_0^\Lambda dp\ p^2 \bigg(\frac{\tanh \left(\frac{1}{2} \beta  \omega_-(p,\Delta) \right)}{\omega_-(p,\Delta)}+\frac{\tanh \left(\frac{1}{2} \beta  \omega_+(p,\Delta) \right)}{\omega_+(p,\Delta)}\bigg)\ .
\label{gapBCS}
\ee
This gives the usual critical curve for a second-order phase transition for the order parameter $\Delta $
as a function of the temperature.
There are  standard approximations that one can do. The second term inside the integrand comes from the antiparticle and can be neglected as $\omega_-\ll \omega_+$ near the Fermi surface. In doing so one connects with the expressions of the non-relativistic case. Near the Fermi surface one may also approximate the factor $p^2$ in the numerator by $\mu^2-m^2$.


\section{Supersymmetric BCS}

Let us now try to design a supersymmetric Lagrangian which incorporates these basic features.
We are interested in a supersymmetric theory with a global baryonic $U(1)_B$ or $U(1)_R$ symmetry which undergoes spontaneous symmetry breaking.
In general, $U(1)$ symmetry breaking is easy to achieve  by a suitable choice of the superpotential $W$.
However, here we are looking for a BCS type mechanism, where the breaking is caused by fermion condensation triggered by quantum effects.
$\mathcal N=1$ supersymmetric models with a canonical K\" ahler potential do not contain any quartic fermion interaction for any choice of superpotential $W$ (see section 4). Quartic fermion interactions arise
 by means of the following choice of  K\" ahler potential:
\be
K(\Phi,\Phi^\dagger ) = \Phi^\dagger \Phi + g^2 (\Phi^\dagger \Phi )^2\ .
\ee
{} We would like to construct a supersymmetric BCS theory with Dirac fermions, and in $\mathcal N=1$ supersymmetric theories 
this requires at least two chiral superfields (a single chiral superfield describes a Weyl fermion).
The simplest theory consists on  two chiral superfields $X$ and $Y$ with the  K\" ahler potential
\be
K(X,Y, X^\dagger, Y^\dagger ) =X^\dagger X +Y^\dagger Y + g^2 (X^\dagger X )^2+ g^2 (Y^\dagger Y )^2\ .
\label{kal}
\ee
The coupling $g$ could in principle be different for the interaction terms involving $X$ and $Y$ superfields. One could
also add, for example, a term $X^\dagger X Y^\dagger Y $ (used in \cite{Buchmuller:1982ty}).
However we shall consider  the above simple choice which already illustrates the essential points. 


\subsection{Chemical potential for $U(1)_B$}\label{secU(1)B}

We first consider the $\mathcal N=1$ supersymmetric model defined in terms of two chiral superfields with  K\" ahler potential (\ref{kal})
and superpotential:
\be
W= m XY\ .
\ee
This gives  masses to scalars and fermions. 
It will be shown that this model is {\it not} suitable to implement BCS mechanism
in supersymmetric theories. The model will illustrate the typical problems that one has to deal with.

We  first consider the Lorentzian theory on ${\bf R}^4$. For the finite temperature theory,
we shall later consider the Euclidean theory on $S^1\times {\bf R}^3$, and eventually on $S^1\times S^3$.
In components, the Lagrangian reads 
\begin{align}
\mathcal L_S&=(1+4g^2\vert\phi_x\vert^2)\partial_\mu\phi_x^*\partial^\mu\phi_x
-\frac{m^2\vert\phi_y\vert^2}{1+4g^{2}\vert\phi_x\vert^2}+(x\leftrightarrow
y)\\
\mathcal L_F&=i(1+4g^2\vert\phi_x\vert^2)(\psi^\dagger_x\bar\sigma^\mu\partial_\mu\psi_x)+4ig^2(\psi^\dagger_x\bar\sigma^\mu\psi_x)\phi_x^*\partial_\mu\phi_x+\frac{g^{2}(\psi_x\psi_x)(\psi^\dagger_x\psi^\dagger_x)}{1+4g^{2}\vert\phi_x\vert^2}\nonumber\\
&\phantom{=\ }+\left(\frac{2mg^2\phi_y\phi_x^*}{1+4g^2\vert\phi_x\vert^2}(\psi_x\psi_x)-\frac{1}{2}m\psi_x\psi_y+h.c.\right)+(x\leftrightarrow
y)\ .
\end{align}
Here we use 
$\eta_{\mu\nu}={\rm diag}(1,\,-1,\,-1,\,-1) $ and as usual
$\bar{\sigma}^{\mu,\, \dot\alpha\alpha}=(\mathbf{1},\,-\sigma^i)$, where $\sigma^i$ are the  Pauli matrices
(we will follow the notation of \cite{terning}).

Note the presence of the (non-renormalizable) quartic fermion interaction. The coupling constant $g$ has dimension of length.
The choice of sign of $g^2$ was made in order to have the same type of interaction as in BCS.
We have checked that the opposite sign does not lead to fermion condensation by quantum effects.
For $g^2<0$  there is no
consistent solution to the gap equation for the vacuum condensate. The effective potential is unstable and
cannot be consistently minimized in the one-loop approximation.
Therefore in what follows we assume $g^2>0$.

We need to introduce a chemical potential and consistency demands that this is
coupled to a conserved non-anomalous $U(1)$ current. 
The superfields $X$ and $Y$ carry opposite $U(1)$ charge so the baryonic $U(1)_B$ current is non-anomalous.
Turning on a chemical potential corresponds to turning on a background $U(1)_B$ gauge field component
$A_0=\mu$.
In order to have a Lagrangian quadratic in fermion fields, one can make a  Hubbard-Stratonovich
transformation in the component Lagrangian by 
introducing two auxiliary fields, $\Delta_x, \Delta_y$, 
\begin{align}\label{LHS}
\mathcal L_S&=(1+4g^2\vert\phi_x\vert^2)D_\mu\phi_x^*D^\mu\phi_x
-\frac{m^2\vert\phi_y\vert^2}{1+4g^{2}\vert\phi_x\vert^2}-g^2(1+4g^{2}\vert\phi_x\vert^2)\vert\Delta_x\vert^2+(x\leftrightarrow
y)\\
\label{LHS2}
\mathcal L_F&=i(1+4g^2\vert\phi_x\vert^2)(\psi^\dagger_x\bar\sigma^\mu D_\mu\psi_x)+4ig^2\phi_x^*D_\mu\phi_x(\psi^\dagger_x\bar\sigma^\mu\psi_x)\nonumber\\
&\phantom{=\ }+\left(\left(\frac{2mg^2\phi_x^*\phi_y}{1+4g^2\vert\phi_x\vert^2}+g^2\Delta_x\right)(\psi_x\psi_x)-\frac{1}{2}m\psi_x\psi_y+h.c.\right)+(x\leftrightarrow
y)\ ,
\end{align}
where $D_\nu=\partial_\nu - iq  \mu \delta_{\nu 0}$ (with no loss of generality one can set the $X$ $U(1)$ charge $q_X=1$, as it can be absorbed into a redefinition of $\mu$; in this way $Y$ has charge $q_Y=-1$).
The Lagrangian has now become quadratic  in the fermion fields,  no quartic fermion interaction is left.
As a result, the functional integral over fermions can be directly performed.

Next, we expand the scalar fields around their VEV's, $\phi=v+\varphi$,
and retain only up to quadratic terms in the scalar fields (we assume real $v$).
We find 
\begin{align}
\mathcal L_S&=(1+4g^2v_x^2)\partial_\mu\varphi_x^*\partial^\mu\varphi_x+4 g^2v_x^2 \left(\mu ^2-\frac{4 g^2 m^2v_y^2}{\left(1+4 g^2 v_x^2\right)^3}\right) (\varphi_x^2+\varphi_x^{*2})
\label{LBUBflat}\\
&\phantom{=\ }
+\frac{4 g^2 m^2 v_x v_y}{\left(1+4 g^2 v_x^2\right)^2}(\varphi_x\varphi_y+\varphi_x^*\varphi_y+\varphi_x\varphi_y^*+\varphi_x^*\varphi_y^*)\nonumber\\
&\phantom{=\ }
+\left(\left(1+16 g^2  v_x^2\right) \mu ^2-4g^{4}\vert\Delta_x\vert^2-\frac{4 g^2 m^2 \left(-1+4 g^2v_x^2\right) v_y^2}{\left(1+4 g^2 v_x^2\right)^3}\right) \vert\varphi_x\vert^2-\frac{m^2}{1+4 g^2 v_x^2} \vert\varphi_y\vert^2\nonumber\\
&\phantom{=\ }
+i\mu(1+8g^2v_x^2)(\varphi_x^*\partial_t\varphi_x-\varphi_x\partial_t\varphi_x^*)-4i\mu
g^2v_x^2(\varphi_x^*\partial_t\varphi_x^*-\varphi_x\partial_t\varphi_x)+(x\leftrightarrow
y,\,\mu\rightarrow-\mu)\\
\mathcal L_F&=i(1+4g^2v_{x}^2)(\psi^\dagger_x\bar\sigma^\mu \partial_\mu\psi_x)+\mu(1+8g^2v_{x}^2)(\psi^\dagger_x\bar\sigma^0\psi_x)\nonumber\\
&\phantom{=\ }+\left(\left(\frac{2mg^2v_xv_y}{1+4g^2v_{x}^2}+g^2\Delta_x\right)(\psi_x\psi_x)-\frac{1}{2}m\psi_x\psi_y+h.c.\right)+(x\leftrightarrow
y,\,\mu\rightarrow-\mu)
\label{LFUBflat}
\end{align}
with 
\be
V_{\rm cl}= \frac{m^2v_y^2}{1+4g^2v_x^2}+(1+4g^2v_x^2)(g^2\vert\Delta_x\vert^2-\mu^2v_x^2)+(x\leftrightarrow
y)\ .
\label{clasV}
\ee
To have canonically normalized kinetic terms, one can redefine fields as follows:
\begin{equation}
\varphi\rightarrow\frac{\varphi}{\sqrt{1+4g^2  v^2}}\ ,
\qquad
\psi\rightarrow\frac{\psi }{\sqrt{1+4g^2  v^2 }}\ .
\end{equation}
Integrating over $\psi,\psi^\dagger, \varphi,\varphi^*$ leads to a one-loop potential depending on $g, v, \Delta, \mu, m$.
Since the model is not renormalizable (just like BCS)  integrals will be regularized  by a momentum cutoff,
representing a ``Debye" energy where new microscopic physics appears.

\medskip

We proceed as follows.
Calling $O_S$ and $O_F$ to the resulting $4\times 4$ scalar and fermion matrices for the quadratic terms in momentum space, we shall write the determinants as:
\be
\det O_S  = \prod_{i=1}^4 \big( \omega^ 2- \omega_{Si}^ 2 \big)\ ,\qquad \det O_F  = \prod_{i=1}^4 \big( \omega^ 2- \omega_{Fi}^ 2 \big)\ ,
\label{eig}
\ee
where 
\be
\omega_{Si}=\omega_{Si}(\mu,|\vec p|, g,m, v_x,v_y, \Delta_x,\Delta_y )\ ,\qquad \omega_{Fi}=\omega_{Fi}(\mu,|\vec p|, g,m, v_x,v_y, \Delta_x,\Delta_y )\ .
\ee
The expressions for  $O_S$ and $O_F$ are shown in appendix B.
The eigenvalues for the frequencies have complicated expressions when $v_x$ and $v_y$ are non-vanishing.
The strategy is to look for non-trivial minima at $v_x=v_y=0$ with $\Delta_x,\ \Delta_y\neq 0$, assuming them to be real. 
Next, we shall check that the one-loop effective potential is locally stable in $v_x$ and $v_y$ directions,
a property that will be ensured by the presence of a mass term.

When $v_x=v_y=0$
the scalar and fermion quadratic terms greatly simplify. At this point, we find the following eigenvalues for the frequency. 
\bea
&& \omega_{S\ 1,2} =  \sqrt{4 g^4 \Delta _x^2+m^2+p^2} \pm  \mu\ ,
\nonumber\\
&& \omega_{S\ 3,4} = \sqrt{4 g^4 \Delta _y^2+m^2+p^2}  \pm \mu\ ,
\eea
\bea
&&\omega_{F\ 1,2}^2 =
2 g^4 \Delta _x^2+2 g^4 \Delta _y^2  +\mu ^2+m^2+p^2\pm {\cal E}_+,\ 
   \nonumber\\
&&\omega_{F\ 3,4}^2 =     2 g^4 \Delta _x^2+2 g^4 \Delta _y^2+\mu ^2+m^2+p^2
\pm {\cal E}_-,
\eea
\be
{\cal E}_\pm =2\sqrt{\mu ^2
   \left(m^2+p^2\right)+g^8 \left(\Delta _x^2-\Delta _y^2\right){}^2
+g^4\left(m^2 \left(\Delta _x+\Delta _y\right)^2\pm 2 \mu  p \left(\Delta _x^2-\Delta _y^2\right)\right)}
\ee

{} For configurations with $\Delta_x=\Delta_y\equiv \Delta$,
the fermion frequencies become 
\be
\omega_F=\sqrt{\left(\sqrt{p^2+m^2+4g^4\Delta^2 {m^2\over \mu^2}}\pm \mu \right)^2 + 4g^4\Delta^2\left( 1-{m^2\over\mu^2}\right)}\ \ .
\ee
On the other hand, for  $\Delta_x=-\Delta_y\equiv \Delta$, we find
\be
\omega_F=\sqrt{\big(\sqrt{p^2+m^2}\pm \mu \big)^2 + 4g^4\Delta^2}\ .
\ee
This is the  same dispersion relation as in the relativistic BCS system of section 2.
This might suggest that BCS mechanism can be implemented in a similar way.
But the presence of charged scalars demands some care.
We first need to identify the Fermi surfaces. 
For $\Delta_x =\Delta_y=0$, they lie on the region where $\omega_{F\ 2,4}$ vanish,
i.e. at
\be\label{FSflat}
\sqrt{p_F^2+m^2}=\mu \ .
\ee
As in the standard relativistic BCS case, the existence of a Fermi surface would require $\mu > m$.
However, in the present supersymmetric system we cannot set $\mu > m$ 
because the scalar contribution to the thermal partition function
\be
{1\over\beta} \sum_i \int \frac{d^3p}{(2\pi)^3} \log \big( 1 -  e^{ - \beta \omega_{Si}} \big)\ ,\qquad 
\ee
is ill-defined, because $\omega_{S\ 2,4}$ become negative below some momentum. The system presents BE condensation, the occupation number of scalars with zero momentum goes to infinity as $\mu $ approaches $m$ from below.
This spoils the BCS mechanism.

One possible approach to elude this problem while maintaining supersymmetry is to
 put the theory on $S^1\times S^3$. Because the scalar field couples to the curvature 
(see e.g. \cite{Festuccia:2011ws}), this will provide an extra mass term for the scalar fields, which might
 allow for  regions in parameter space with Fermi surfaces, and without problems of BE condensation.
The mass term, when the $R$-charge of the scalars is one, is now of the form
$$
\left(m^2+ R^{-2}\right) (\phi_x^* \phi_x+\phi_y^*\phi_y)\ ,
$$
where $R$ is the radius of the three-sphere.
The scalar contribution would be negligible if one could assume that $1/R>\Lambda $.
However,
having put the theory on  $S^3$,  the integral over momentum is replaced by a discrete sum originating from the Kaluza-Klein modes of $S^ 3$.
This replacement is achieved by 
\bea
&&{\rm Scalars:}\qquad  p^2\ \longrightarrow \  l (l+2)R^{-2} 
\nn\\
&&{\rm Fermions:}\ \ \  p^2\ \longrightarrow \ (l+1/2)^2 R^{-2}\label{p->l}
\eea
with $l=0,1,2,\ldots$.  
One must also take into account the degeneracy: for scalars, $d_l^S= (l+1)^ 2$; for fermions,  $d_l^F= l(l+1)$. In particular, for the fermions,  $l=0$ does not contribute.
{} For the scalars, in addition we must add the mass term $R^{-2}$.
This is effectively incorporated by the replacement
\be
 p^2\ \longrightarrow \  l (l+2)R^{-2} + R^{-2}=(l+1)^ 2 R^{-2}
\ee
These formulas show that one cannot assume $1/R>\Lambda $, since such cutoff  would leave no excitation in the system.
Therefore it is not possible to separate the scalar mass scale from the Fermi surface.

In order to see if the system can have Fermi surfaces, we need the detailed form of the Lagrangian
on $S^3$. This depends on the $R$ charges of the fields. We denote by $q$ the $R$ charge of $\phi_x$ so that the charge of $\phi_y$ is $2-q$.
From the expressions given in \cite{Festuccia:2011ws}, we find
\begin{align}
\mathcal L_S&=(1+4g^2\vert\phi_x\vert^2)\partial_\mu\phi_x^*\partial^\mu\phi_x\nonumber\\
&\phantom{=\ }
+\left(\frac{q(q-2)}{R^2}+2\mu\frac{q-1}{R}+\mu^2\right)\vert\phi_x\vert^2+4g^2\left(\frac{q(q-1)}{R^2}+\mu\frac{2q-1}{R}+\mu^2\right)\vert\phi_x\vert^4\nonumber\\
&\phantom{=\ }
+i\left(\frac{q-1}{R}+\mu\right)(\phi_x^*\partial_t\phi_x-\phi_x\partial_t\phi_x^*)+2ig^2\left(\frac{2q-1}{R}+2\mu\right)\vert\phi_x\vert^2(\phi_x^*\partial_t\phi_x-\phi_x\partial_t\phi_x^*)\nonumber\\
&\phantom{=\ }
-\frac{m^{2}\vert\phi_y\vert^2}{1+4g^2\vert\phi_x\vert^2}-g^2(1+4g^2\vert\phi_x\vert^2)\vert\Delta_x\vert^2+(x\leftrightarrow
y,\mu\rightarrow-\mu,q\rightarrow2-q)\\
\mathcal L_F&=i(1+4g^2\vert\phi_x\vert^2)(\psi^\dagger_x\bar\sigma^\mu\partial_\mu\psi_x)+4ig^2(\phi_x^*\partial_\mu\phi_x)(\psi^\dagger_x\bar\sigma^\mu\psi_x)\nonumber\\
&\phantom{=\ }
+\left(\frac{2q-1}{2R}+\mu\right)(\psi^\dagger_x\bar\sigma^0\psi_x)+4g^2\left(\frac{4q-1}{2R}+2\mu\right)\vert\phi_x\vert^2(\psi^\dagger_x\bar\sigma^0\psi_x)\nonumber\\
&\phantom{=\ }
+\left(\frac{2mg^2\phi_x^*\phi_y}{1+4g^2\vert\phi_x\vert^2}+g^2\Delta_x\right)(\psi_x\psi_x)+\left(\frac{2mg^2\phi_x\phi_y^*}{1+4g^2\vert\phi_x\vert^2}+g^2\Delta_x^*\right)(\psi^\dagger_x\psi^\dagger_x)\nonumber\\
&\phantom{=\ }
-\frac{1}{2}m(\psi_x\psi_y+\psi^\dagger_x\psi^\dagger_y)+(x\leftrightarrow
y,\mu\rightarrow-\mu,q\rightarrow2-q)
\end{align}

We shall demand that  in  the unbroken phase the theory has well-defined thermodynamical potentials.
So we begin by  considering the case $\Delta_x=\Delta_y=0$, $v_x=v_y=0$. We will now see that BE condensation is inevitable in this case, which is sufficient to
rule out the model. Consider first the case $q=1$, i.e. the $U(1)$ charges of $X$ and $Y$ are equal to $1$.
The scalar contribution is now given in terms of the frequencies
\be
 \omega_{S} =  \sqrt{(l+1)^2 R^{-2} +m^2} \pm  \mu\ ,\qquad l=0,1,2,\ldots
\ee
{} If both $X$ and $Y$ had the same baryon charge,  the
Fermi surface would just be determined by  the replacement \eqref{p->l}  in the flat expression \eqref{FSflat},
and shifting the chemical potential by $\mu\rightarrow\mu+1/(2R)$. 
As  $X$ and $Y$
have opposite baryon charges, this is more involved. By  explicitly computing $\omega_F$ from
the above Lagrangian, we obtain that the Fermi surface $\omega_F=0 $ is at 
\begin{equation}
\sqrt{ l_F^2 R^{-2}+m^2}=\mu ,\qquad l_F= 1,2,\ldots
\label{mura}
\end{equation}
{} For a given choice of $l_F$, one can determine $\mu$. Substituting $\mu $ in the  lowest ($l=0$)
scalar frequency, we see that the scalar frequency cannot be positive as
long as $l_F=1,2,\ldots$,
\begin{equation}
 \sqrt{ R^{-2} +m^2} - \sqrt{ l_F^2 R^{-2}+m^2}\leq 0\, .
\end{equation}
Therefore,  even on $S^3$,  it is not possible to separate the Fermi surface
from the region of BE condensation. 
The underlying reason being that the extra mass term for the scalar provided by the coupling to the curvature
of the space
is of the same order as the quantized fermion momentum values. 
The same problem arises for any choice of $q$.




\subsection{A simple supersymmetric BCS model: \\ Chemical potential for $U(1)_R$}\label{secU(1)R}

Let us now consider an $\mathcal N=1$ supersymmetric model with two chiral superfields $X$ and $Y$ with K\"ahler potential given by (\ref{kal}) and superpotential
$W=0$. 
The Lagrangian has a $U(1)_R$ symmetry for arbitrary $U(1)_R$ charges of the $X$ and $Y$ superfields.
It is convenient to consider the $U(1)_R$ symmetry under which scalars $\phi_x$ and $\phi_y$ are neutral, 
so that fermions $\psi_x$ and $\psi_y$  have charge $ -1$.
The advantage of this choice is that we can avoid problems of BE condensation even in ${\bf R}^4$.
Note that with this charge assignation the $U(1)_R$ symmetry is anomalous. However, this can be easily cured by adding to the theory free superfields with canonical
K\" ahler potential with the required $U(1)_R$ charges to cancel the anomaly.
For example, one may add $Z_i$, $i=1,2$ with $R$-charges $R(Z_i)=2$ so that  $\psi_{Z_1},\ \psi_{Z_2}$ have charges $+1$.
The scalars in $Z_i$ would then couple to the chemical potential and may undergo Bose-Einstein condensation. However, this sector is completely decoupled and therefore does not participate in the thermodynamics governing the  $X$, $Y$ sector.

The component Lagrangian with chemical potential included can be obtained from the previous case, (\ref{LHS}), (\ref{LHS2}), 
by setting $m=0$, vanishing $U(1)$ charges for the scalar fields
(which amounts to replace covariant derivatives of the scalar fields by ordinary derivatives) and taking into account that fermions $\psi_x$ and $\psi_y$ now have
the same charge $-1$.
The quadratic Lagrangian for the fluctuations (after expanding around expectation values)
is given by
\begin{align}
\mathcal L_S &=\partial_\mu\varphi_x^* \partial^\mu \varphi_x+\partial_\mu\varphi_y^*\partial^\mu\varphi_y
-\frac{4g^4\vert\Delta_x\vert^2}{1+4g^2v_x^2}
\vert\varphi_x\vert^2-\frac{4g^4\vert\Delta_y\vert^2}{1+4g^2 v_y^2}\vert\varphi_y\vert^2\ ,
\label{LBUR}\\
\mathcal   L_F &=i(\psi^\dagger_x\bar\sigma^\mu\partial_\mu\psi_x)+
i(\psi^\dagger_y\bar\sigma^\mu\partial_\mu\psi_y)
-\mu(\psi^\dagger_x\bar\sigma^0\psi_x)-\mu(\psi^\dagger_y\bar\sigma^0\psi_y)\nonumber\\
&\phantom{=\ }+\left(\frac{g^2\Delta_x}{1+4g^2v_x^2}(\psi_x\psi_x)+ \frac{g^2\Delta_y}{1+4g^2v_y^2}(\psi_y\psi_y)+h.c.  \right)\ ,
\label{LFUR}
\end{align}
where we have rescaled the fields to have canonical kinetic terms.
The classical potential is given by
\be
V_{\rm cl}= g^2 \left(4 g^2 v_x^2+1\right)| \Delta_x|^2+(x\leftrightarrow
y)\ .
\label{clV}
\ee
The equations of motion for $\Delta_x,\ \Delta_y$ give (setting the scalar fluctuations $\varphi_x, \varphi_y\to 0$)
\be
\Delta_x=\frac{\psi^\dagger_x\psi^\dagger_x}{(1+4g^2 v_x^2)^2}\ ,\qquad \Delta_y=\frac{\psi^\dagger_y\psi^\dagger_y}{(1+4g^2 v_y^2)^2}\ .
\ee
$\Delta_x,\ \Delta_y$ have both $U(1)_R$ charges equal to 2. Vacuum expectation values for them thus spontaneously break $U(1)_R$
and represent a measure of the fermion condensate.

By proceeding in a similar way as in the previous case, we now find the following frequencies for scalars and fermions
\begin{align}
\omega_{S\ 1,2}^2&=p^2+\frac{4 g^4 \Delta _x^2}{1+4 g^2 v_x^2}
\ ,&&  \omega_{S\ 3,4}^2=p^2+\frac{4 g^4 \Delta _y^2}{1+4 g^2 v_y^2}\ ,
\\
\omega_{F\ 1,2}^2&=(p\pm\mu )^2+\frac{4 g^4 \Delta _x^2}{\left(1+4 g^2 v_x^2\right)^2}\ , && 
\omega_{F\ 3,4}^2=(p\pm\mu )^2+\frac{4 g^4 \Delta _y^2}{\left(1+4 g^2 v_y^2\right)^2}\ .
\end{align}
Here we have chosen real $\Delta_x,\ \Delta_y$, as one-loop potential depends only on their moduli.
We stress that these simple dispersion relations are a consequence of the 
extreme simplicity of this supersymmetric model; generic models (even with simple superpotentials) typically lead to very complicated eigenvalues for the frequencies. 

In the present case, the dynamics of the $X$ and $Y$ fields are  decoupled. 
It is clear that the same configuration that minimizes the one-loop potential in the $x$ direction also minimizes
the one-loop potential in the $y$ direction. Therefore with no loss of generality 
we set $v_x=v_y\equiv v$ and $\Delta_x=\Delta_y\equiv \Delta$.

The complete one-loop thermodynamic potential  is given by
\begin{align}
\Omega&=2 g^2 \left(1+4 g^2 v^2\right) \Delta^2+\frac{1}{\pi^2\beta}\int_0^\Lambda dp\ p^2\left(2\log \left[\sinh \frac{\beta}{2}\sqrt{p^2+\frac{4g^4\Delta ^2}{1+4 g^2 v^2}}\right]\right.\nonumber\\
&\phantom{=\ }
-\log\left[ \cosh\frac{\beta }{2}\sqrt{(p+\mu )^2+\frac{4g^4\Delta ^2}{\left(1+4 g^2 v^2\right)^2}}\right]
\left.-\log\left[ \cosh\frac{\beta }{2}\sqrt{(p-\mu )^2+\frac{4g^4\Delta ^2}{\left(1+4 g^2 v^2\right)^2}}\right]\right)\ .
\end{align} 
When the vacuum lies at $\Delta\neq 0$, then $v=0$ is a local minimum. When the vacuum lies at $\Delta=0$, then there is a flat direction in  $v$, because in this case the frequencies do not depend on $v$. This  is confirmed by the evaluation of the one-loop potential.

The gap equation  $\Delta =\Delta(T)$ is determined by the equation
\be
{d\Omega\over d\Delta}=0\ .
\ee
This gives, when $v=0$,
\bea
1 &=&\frac{g^2 }{2\pi ^2}\int_0^\Lambda dp\ p^2 \bigg(\frac{\tanh \left(\frac{1}{2} \beta  \sqrt{4 g^4 \Delta ^2+(p-\mu )^2}\right)}{\sqrt{4 g^4 \Delta ^2 +(p-\mu )^2}}+\frac{\tanh \left(\frac{1}{2} \beta  \sqrt{4 g^4 \Delta ^2 +(p+\mu )^2}\right)}{\sqrt{4 g^4 \Delta ^2 +(p+\mu )^2}}
\nonumber\\
&&\hspace{3cm}- \frac{2 \coth \left(\frac{1}{2} \beta  \sqrt{4 g^4 \Delta ^2 +p^2}\right)}{\sqrt{4 g^4 \Delta ^2 +p^2}}\bigg)\ .
\label{gappa}
\eea
The gap equation can be compared with the gap equation (\ref{gapBCS}) of the relativistic BCS system of section 2.
One difference is that now scalars and fermions have zero mass, since
a mass term $mXY$ would not be consistent with scalars neutral under $U(1)_R$.
The second and more fundamental difference is given by the scalar contribution --represented by the second line in the above equation-- that we analyze in what follows.

An important consequence  of supersymmetry is that the critical curve $\Delta (T)$ now depends logarithmically with the cutoff:
for large $p$, the integral in (\ref{gappa}) behaves as follows:
\be
\frac{g^2}{\pi^2}\int^\Lambda dp\ {\frac{ \mu^2}{p}} \sim \log\Lambda\ .
\ee
If the scalar contribution is removed, like in non-supersymmetric BCS, one has instead
\be
\frac{g^2}{\pi^2}\int^\Lambda dp\ { p} \sim \Lambda^2\ .
\ee
Obviously, a logarithmic dependence with the UV cutoff is a desirable feature, since the thermodynamics becomes much less sensitive to the underlying microscopic physics.

At the same time, the IR physics produced by the scalar sector has a striking effect:
the superconducting transition becomes first-order, instead of  second-order, as it would be in standard BCS.
The IR physics of the scalar sector is important at the onset of the transition, where $\Delta $ is small.
To see the nature of the transition, we need to compute $d\Delta/dT$. This can be obtained by differentiating the gap equation $d\Omega/ d\Delta$
with respect to $T$. Writing the gap equation in the form
$1=f(\Delta^2,T)$,
one has 
\be
{d\Delta\over dT} = -  \frac{  \frac{\p^2 \Omega} {\p T\p \Delta} }{ \frac{\p^2 \Omega}{\p \Delta^2}   }= -  \frac{1}{2\Delta }\ \frac{  \frac{\p f} {\p T} }{\frac{\p f}{\p (\Delta^2)}   }\ .
\ee
In a second-order phase transition, $d\Delta/ dT $ is singular at the critical temperature, where 
$\Delta =0$.
This is because $\p f/\p T$ and $\p f/\p (\Delta^2) $ are regular at $\Delta =0$.
While the scalar contribution to the one-loop potential is  regular at $\Delta =0$, its second derivative with respect to $\Delta^2 $
has a singularity near $\Delta =0$ originating from the region near $p=0$. We have
\be
\frac{\p f}{\p (\Delta^2)}   \approx   \frac{8 g^{6}T}{\pi ^2 }\int_0 dp\ p^2 \frac{1}{(4 \Delta ^2 g^4+p^2)^2 }
\approx    \frac{g^{4}T}{\pi  }\ \frac{1}{ \Delta  }\ .
\ee
As a result, $d\Delta/ dT$ is now finite at $\Delta =0$.
The superconducting phase transition is therefore first-order. This significant change coming from the $p=0$ region would obviously not take place
if the scalar field was massive. In such a  case, the phase transition would still be second-order. But, as explained, in the present model it is not possible to add a mass term.

\begin{figure}[h!]
\hspace*{-11mm}
\includegraphics[width=.35\textwidth]{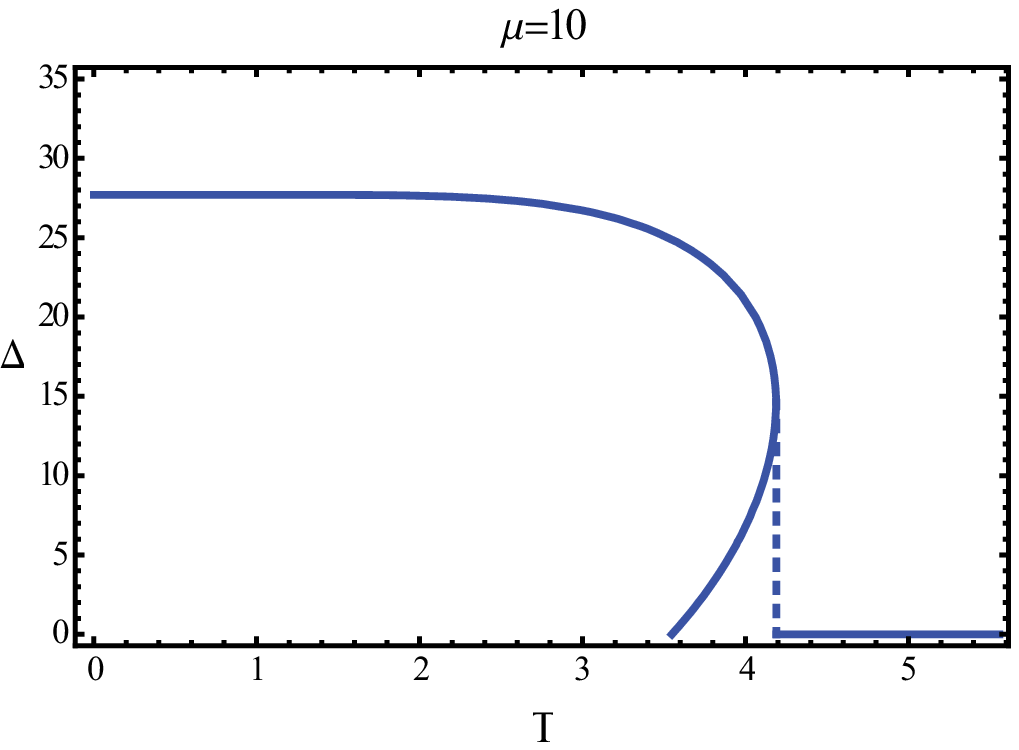}
\includegraphics[width=.345\textwidth]{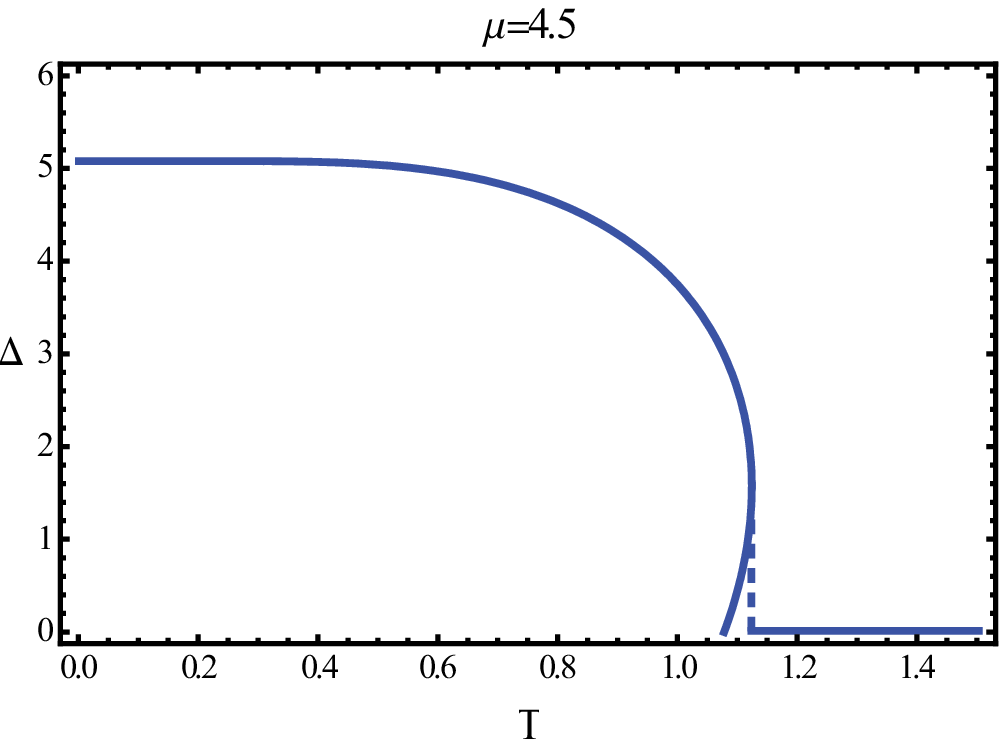}
\includegraphics[width=.35\textwidth]{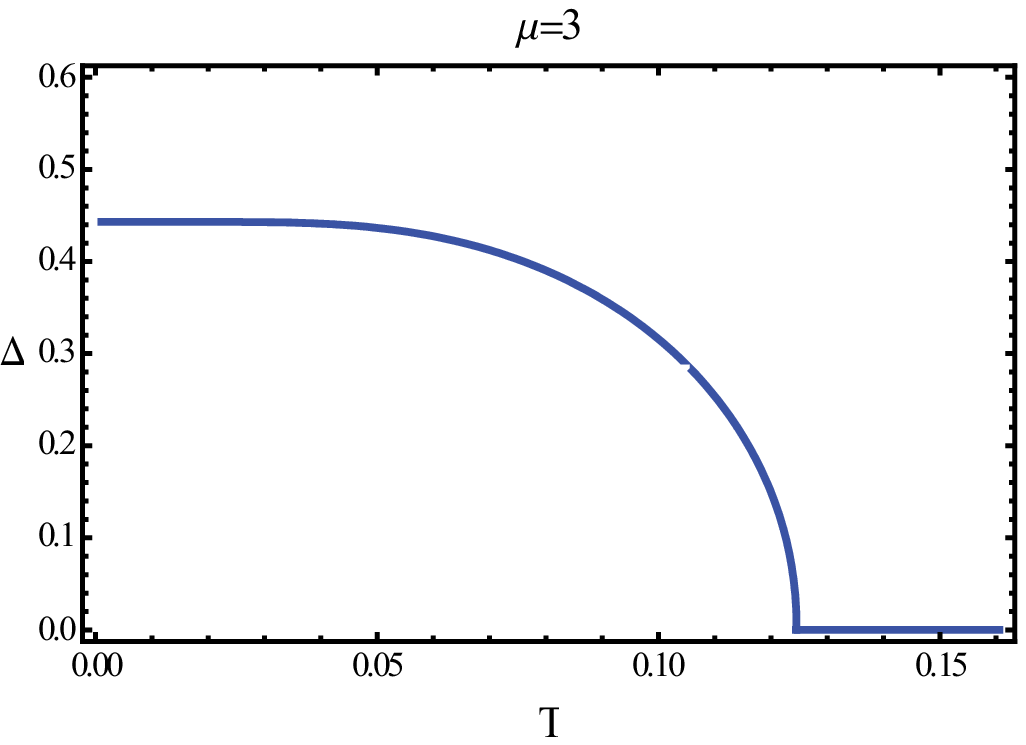}
\caption{Gap $\Delta $ vs. $T$ at  $\mu =10,\ 4.5,\ 3 $ ($g=0.5$, $\Lambda=40$). }
\label{Gap}
\end{figure}

$\Delta=\Delta(T)$ is shown
in fig.~\ref{Gap}  for different values of the chemical potential. We see that, as the chemical potential gets smaller, the transition approaches to a
second-order phase transition.
In general, the scalar field has the effect of decreasing the critical temperature with respect to the relativistic BCS case.

\begin{figure}[h!]
\centering
\includegraphics[width=.5\textwidth]{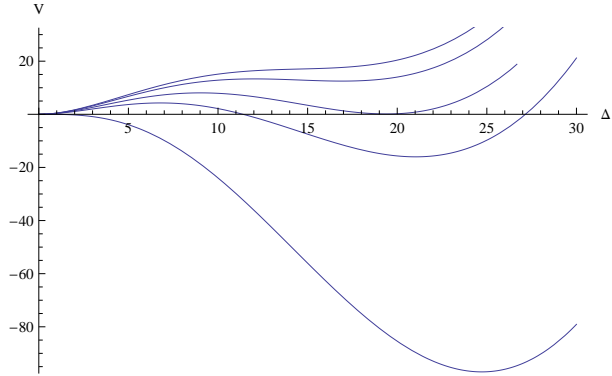} 
\caption{Potential showing generation of minimum at $\Delta\neq 0$. Different curves (from top to bottom) 
at  $T=4.2,\ 4.17,\ 4.09, \ 4,\ 3.55$ (with $g=0.5$, $\mu=10$, $\Lambda=40$).}
\label{Pot}
\end{figure}

{} In an interval of temperature, $T_{c1}<T<T_{c2}$,  there are three branches --characteristic of first-order phase transitions--
corresponding to three solutions of the gap equation: the trivial minimum at $\Delta =0$, a maximum and another minimum at higher $\Delta $.
These are exhibited in fig.~\ref{Pot}, showing the one-loop potential at different temperatures. We see how the non-trivial maximum and minimum are created
as the temperature is lowered below a certain critical value $T_{c2}$ ($T_{c2}\approx 4.2$ in fig.~\ref{Pot}; see also fig.~\ref{Gap} with $\mu=10$). 
At a temperature $T_c$,  the non-trivial minimum becomes degenerate with
the minimum at $\Delta=0$ (in  fig.~\ref{Pot}, this occurs at $T_c\approx 4.09$). 
In the interval $T_{c1}<T<T_{c}$, the symmetric vacuum $\Delta =0$ is metastable.
Below $T_{c1}\approx 3.55$,  the  symmetric vacuum $\Delta =0$ becomes unstable and the only minimum of the potential is the
SSB vacuum at $\Delta\neq 0$.


The gap at $T\to 0$ can be determined analytically. In this limit, the hyperbolic tanh and coth become
1 and the integrals can be easily computed. Assuming $\Lambda\gg g\Delta ,\ \mu$,
we find the following result
\be
1\approx \frac{g^2 \mu ^2}{2
   \pi ^2} \left( 2\log \Big(\frac{\Lambda }{g^2\Delta(0)  }\Big)-3\right)\
   ,
\ee
i.e.
\be
\big|\Delta(0)\big|\approx \frac{\Lambda\  e^{-\frac{\pi ^2}{g^2 \mu ^2}-\frac{3}{2}}}{g^2}\ .
\label{arfe}
\ee
It should be noted that our normalization for $\Delta $ is different from the normalization
of $\Delta $ of section 2. One can easily go back to the normalization of section 2 by the substitution
$2g^2\Delta\to \Delta_{\rm susyBCS}$.
Taking into account this, we are finding
\be
\big|\Delta(0)_{\rm susyBCS}\big| \approx 2\Lambda\  e^{-\frac{\pi ^2}{g^2 \mu ^2}-\frac{3}{2}}\ .
\label{ops}
\ee
This is similar to the standard formula in BCS theory with the identification $\mu^2=\pi^2 N(0)$, where
$N(0)$ represents the electronic density of states at the Fermi energy.
It is useful to reproduce the non-supersymmetric formula in the present context.
We consider the gap equation (\ref{gappa}), dropping the scalar and the antiparticle contribution.
As explained in section 2, the latter is negligible near the Fermi surface.
We obtain the equation
\be
1 =\frac{g^2 }{2\pi ^2}\int_0^\Lambda dp\ p^2 \frac{1}{\sqrt{\Delta(0)_{\rm BCS} ^2 +(p-\mu )^2}}\ .
\label{gapNS}
\ee
This formula diverges quadratically. It does not give a sensible result if the `Debye energy' $\Lambda $ is  far from the Fermi energy.
The standard procedure involves an {\it ad hoc} approximation, where $p^2$ in the numerator is replaced by $\mu^2 $
and the integral is done in the interval $|p-\mu|<\Lambda$.
This leads to the result
\be
\big|\Delta(0)_{\rm BCS}\big| \approx 2\Lambda\ \frac{ e^{-\frac{\pi ^2}{g^2 \mu ^2}} }{1- e^{-\frac{2\pi ^2}{g^2 \mu ^2}}}\ .
\ee
In the supersymmetric case, because  quadratic divergences are canceled between fermion and scalar contributions, the analog result (\ref{ops}) follows by performing the integrals in the gap equation exactly, with no need of  the {\it ad hoc} approximation $p^2\to \mu^2 $.


\medskip

Let us now study the specific heat. It is instructive to examine the different contributions to the thermodynamic potential closely. 
Consider first the symmetric phase $\Delta=0$.
As we shall be interested in derivatives with respect to the temperature,
we can subtract the Coleman-Weinberg contribution, so that integrals are convergent.
We write
\be
\Omega =\Omega_{\rm scalar} + \Omega_{\rm electron}+\Omega_{\rm positron}\ ,
\ee
where (after integration by parts)
\be
\Omega_{\rm scalar}\big|_{\Delta=0} =-\frac{2}{3\pi^2}\int_0^\infty dp\ p^3\ \frac{e^{-\frac{p}{T}}}{1-e^{-\frac{p}{T}}}\ ,
\ee
\be
\Omega_{\rm electron}\big|_{\Delta=0} =-\frac{1}{3\pi^2}\int_0^\infty dp\ p^3\  \frac{e^{-\frac{p-\mu }{T}}}{1+e^{-\frac{p-\mu }{T}}}\ ,
\ee
\be
\Omega_{\rm positron}\big|_{\Delta=0} =-\frac{1}{3\pi^2}\int_0^\infty dp\ p^3\  \frac{e^{-\frac{p+\mu }{T}}}{1+e^{-\frac{p+\mu }{T}}}\ .
\ee
We get
\be
\Omega_{\rm scalar}\big|_{\Delta=0} =-\frac{2}{45} \pi ^2 T^4\ ,
\ee
as expected, since, 
when $\Delta=0$, the scalar contribution describes a relativistic boson particle
(there is an extra factor of 4 as compared with the usual single scalar contribution, because we have two complex scalar fields $\varphi_x,\ \varphi_y$).
The integrals for the fermion contributions can be computed analytically in terms of polylogarithmic functions.

Let us now compute the different contributions to the specific heat.
The entropy and specific heat are given by the familiar formulas
\be
S= -\left( {\partial \Omega\over \partial T}\right)_\mu \ ,  \qquad c=T  \left({dS\over dT}\right)_\mu \ .
\ee
Note that  specific heat here is defined as a partial derivative at constant $\mu$, instead of constant charge density $\rho $. In the present case, we find this to be a more sensible
quantity, since the scalars are neutral and would not
contribute to the constraint $\rho=d\Omega/d\mu$.

We obtain 
\be
c_{scalar}\big|_{\Delta=0}  =\frac{8 \pi ^2 T^3}{15}\ ,
\ee 
as usual for  relativistic bosons. Consider now the fermion contributions.
At large $T$, the dependence on $\mu$ disappears and one gets the usual behavior  of a relativistic fermion
\bea
c_{electron}\big|_{\Delta=0}  = c_{positron}\big|_{\Delta=0}=\frac{7 \pi ^2 T^3}{30}\ ,\qquad {\rm for}\ \ T\gg\mu\ .
\eea
At low temperatures,  $c_{positron}$ is exponentially suppressed, $c_{positron}\sim e^{-\mu/T}$.
For the electron, the integral  picks the main contribution near the Fermi surface, $p\sim\mu$,
and one gets the usual linear behavior for the electronic specific heat at low temperatures
\be
c=c_{electron}\big|_{\Delta=0} \sim  \ \frac{ \mu^2  T }{3}\ .
\ee

Let us now compute the full $c(T)$ including the region $T<T_{c2}$ where $\Delta\neq 0$.
We use  the notation $\varepsilon =\Delta^2$.
Then
\be
   c(T) = - T  \left( \frac{ \p^2\Omega}{\p T^2} + \frac{\partial ^2\Omega} {\partial T\partial \varepsilon}  \frac{\partial \varepsilon}{\partial T} \right)\ ,\qquad \frac{\partial \varepsilon}{\partial T}  = - \frac{  \frac{\p^2 \Omega} {\p T\p \varepsilon} }{ \frac{\p^2 \Omega}{\p \varepsilon^2}   }\ ,
\ee
where we used the fact that $\Delta(T)$ is defined by  $\partial \Omega/\p \Delta=0$.
The resulting  $c(T)$ is shown in fig. 3. 
At $T<T_{c2}$, it exhibits the expected exponential suppression due to the gap. 
At the critical temperature, we notice  the characteristic discontinuity of first-order phase transitions. 
In first-order phase transitions the entropy may experience a finite jump at the transition, leading to an infinite jump in the specific heat.
For $T_{c2}<T <O(\mu)$, we see the  linear behavior coming from the electron contribution.
Finally, at high temperatures, it exhibits the   $T^3$ behavior  shown above 
in terms of analytic formulas.

\begin{figure}[h!]
\centering
\includegraphics[width=.5\textwidth]{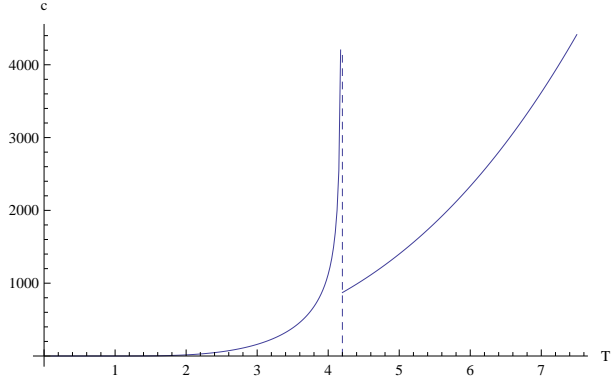} 
\caption{Specific heat as a function of the temperature (with $g=0.5$, $\mu=10$, $\Lambda=40$).}
\label{heatc}
\end{figure}

\bigskip

\section{Models with canonical K\" ahler potential}

 Let us now consider the possibility of implementing a BCS mechanism in models with canonical K\"ahler potential.
Fermion condensation  now needs to be triggered by interactions contained in the superpotential $W$.
A way to find an appropriate model is by starting with the previous model with non-canonical K\"ahler potential
and ``integrating in" some fields. 
One simple example is the model defined by the superpotential
\be
K= X^\dagger X+ Y^\dagger Y\ ,\qquad W= m_0 Z(X- gY^2)\ .
\label{daia}
\ee
The chiral superfield $Z$ has no kinetic term and can be integrated out. The 
equation for $Z$ sets $X=gY^2$.  Substituting this relation
into the K\"ahler potential, one finds a new Lagrangian with $K=Y^\dagger Y+g^2(Y^\dagger Y)^2$,
which contains  quartic fermion interactions, as desired. Thus this model is equivalent to the model considered in section 3
(after the addition of another  superfield similar to $Y$).
The model (\ref{aia}) represents a supersymmetric analog of Hubbard-Stratonovich Lagrangian, with the chiral superfield $X$ playing the role of $\Delta$. Note that there is a $U(1)_B$ as well as a $U(1)_R$ symmetry. One can choose a $U(1)_R$ charge assignment such that $X$ and $Y$ are neutral and $Z$ has charge equal to 2.

Another example is given by a supersymmetric model containing three chiral superfields
$X,\ Y,\ Z$, with potentials
\be
K= X^\dagger X+ Y^\dagger Y\ ,\qquad W= m_0 Z(X- gY^2)+ MZ^2\ .
\label{aia}
\ee
Once again, $Z$ has no kinetic term and can be exactly integrated out.
We end up with the effective model
\be
K= X^\dagger X+ Y^\dagger Y \ ,\qquad W= m (X-gY^2)^2\ ,\ \ \ m=-{m_0^2\over 4M}\ .
\label{baia}
\ee
It should be noted that this model is not equivalent to (\ref{daia}). 
If one ignored the kinetic term for $X$, the equation for $X$ would set $X=gY^2$
(the same result of course follows by working in terms of component fields).
But it is not legitimate to ignore the kinetic term $X^\dagger X$ as this  is precisely
the term that will generate the effective four-fermion interactions and the dynamics that we wish to study.
Therefore, understanding the vacuum  dynamics requires a detailed examination of the model.

In what follows we shall investigate
the  three models  with superpotential (\ref{daia}),  (\ref{aia}) and (\ref{baia}), but now
with canonical K\" ahler potential for all fields $X,\ Y,\ Z$.
The models  (\ref{aia}) and (\ref{baia}) have only $U(1)_R$ symmetry, under which $Z$ and $X$ have charge 1 and $Y$ has charge $1/2$.
In all cases, the $U(1)$  symmetries are anomalous, but this can be cured as before by adding a suitable set of free superfields.
We would like to see if these models can reproduce a BCS mechanism leading to a fermion condensate,
similar to the model of section 3.2, even in the case when all  fields are dynamical.
The basic
Wess-Zumino component Lagrangian for the $\mathcal N=1$ supersymmetric  model with canonical K\" ahler potential reads \cite{terning}
\begin{equation}
\mathcal{L}=\partial_{\mu}\phi_i^{*}\,\partial^{\mu}\phi_i
+i\,\psi_i^{\dagger}\,\bar{\sigma}^{\mu}\,\partial_{\mu}\psi_i-\frac{1}{2}\,\Big((\p_i\p_jW)_\phi \,\psi_i\,\psi_j+(\p_i\p_jW)_\phi^{*}\,\psi_i^{\dagger}\,\psi_j^{\dagger}\Big)-|(\p_i W)_\phi|^2\ .
\label{wzl}
\end{equation}
In particular, the component Lagrangian for (\ref{baia})  is
\begin{align}
\mathcal L &=\partial^\mu\phi^*_x\partial_\mu\phi_x+\partial^\mu\phi^*_y\partial_\mu\phi_y+
i\psi^\dagger_x\bar\sigma^\mu\partial_\mu\psi_x+
i\psi^\dagger_y\bar\sigma^\mu\partial_\mu\psi_y\nonumber\\
&\phantom{=\ }-4m^2(1+4g^{2}\vert\phi_y\vert^2)\vert \phi_x-g\phi_y^2\vert^2\nonumber\\
&\phantom{=\ }+\left( -m\psi_x\psi_x+2mg(\phi_x-3g\phi_y^2)\psi_y\psi_y+4mg\phi_y\psi_x\psi_y+h.c.\right)\
.
\end{align}
Consider now the equation of motion for  $\phi_x^*$. For constant fields, one would have
 \be
2m(1+4g^2\vert\phi_y\vert^2)( \phi_x-g\phi_y^2)=g \psi_y^\dagger\psi^\dagger_y\ .
\ee
This equation connects the fermion bilinear $  \psi^\dagger_y\psi^\dagger_y$ with the expectation value of $\phi_x$, in much the same way
in the BCS Lagrangian the fermion bilinear is connected with $\Delta $. In addition to 
  $\psi_y$, its scalar superpartner $\phi_y$ also appears in the equation.
If the term $X^\dagger X$ were removed from the K\" ahler potential, then the 
equation would just be $\phi_x=g\phi_y^2$. After the functional integral over $\psi,\ \psi^\dagger$ and over the scalar field fluctuations is
carried out, we are left with a one-loop potential depending on $v_x=\langle \phi_x\rangle$ and $v_y=\langle \phi_y\rangle$.
The expectation value  $v_x$ at $v_y=0$ is an order parameter and a measure of the fermion condensate.

 Standard BCS theory involves a Bogoliubov transformation, needed to diagonalize the Hamiltonian, which  leads to the usual concept of quasiparticles. In the current system  (\ref{wzl})  with canonical K\"ahler potential, where $\phi_x\sim   \psi_y^\dagger\psi^\dagger_y$, the Bogoliubov transformation is needed in virtue of the presence of $v_x \psi_y\psi_y$ and
$v_x^* \psi^\dagger_y \psi^\dagger_y $ in the quadratic part of the  Lagrangian. Diagonalizing the corresponding Hamiltonian involves a mixing of creation and annihilation operators.

\bigskip 

We summarize the results of the direct  calculation for  these models.

The functional integral over fermions can be explicitly carried out  as the Lagrangians are quadratic in fermions
(the chiral superfield $X$ represents the supersymmetric analog of the Hubbard-Stratonovich field $\Delta $). 
The scalar fields
are expanded around a vacuum value, e.g. $\phi_x=v_x+\varphi_x,\ \phi_y=v_y+\varphi_y,\ \phi_z=v_z+\varphi_z$. 
The conclusions are as follows:

\begin{itemize}

\item $W=m(X-gY^2)^2$.
The model has only $U(1)_R$ symmetry with  $(q_X,q_Y)= (1,1/2)$.  Chemical potential is introduced for this $U(1)$. Like in the model of section 3.1 with baryonic $U(1)_B$ chemical potential,
scalar particles are charged. As a result,  on ${\bf R^4}$ one cannot have Fermi surfaces because they would overlap with the regions of  BE condensation. 
 Therefore, like in that model, we attempt to study the theory on a 3-sphere of radius $R$ (relevant formulas are given in appendix A).
At $v_y=0$, there are Fermi surfaces at $\mu R=2l_F+1$, $l_F=1,2,...$.
{}For this value of $\mu$, one finds a scalar frequency $\omega_S=(1+l-l_F)/R$, $l=0,1,2,...$.
This shows that it is not possible to have positive definite scalar frequencies for any $l_F$.
Thus the Fermi surface cannot be separated from the region of  BE condensation where the thermal one-loop potential is ill-defined.

\item $W=mZ(X-gY^2)+MZ^2$. This is a renormalizable model and can be viewed as an UV completion of the previous case.
It has a $U(1)_R$ symmetry with uniquely determined $U(1)$ charges, $(q_X,q_Y,q_Z)= (1,1/2,1)$, and a similar IR physics as the model  $W=m(X-gY^2)^2$,
with $M$ playing the role of the UV cutoff
$\Lambda $.


\item $W=mZ(X-gY^2)$. This model has a baryonic $U(1)_B$ as well as $U(1)_R$ symmetry.
If chemical potential is introduced for $U(1)_B$, again we find that the Fermi surfaces cannot be separated from the region of BE condensation
(irrespective of the R-charge assignation).

Consider now a chemical potential coupled to the $U(1)_R$ current.
One can assign charges $(q_X,q_Y,q_Z)= (2-q,1-q/2,q)$.
Unlike the model of section 3.2, now it is not possible to have only neutral scalars.
Nonetheless, for $q=2$, i.e. when $(q_X,q_Y,q_Z)= (0,0,2)$, in the unbroken phase $v_x=v_y=v_z=0$,
it is possible to have a Fermi surface
without BE condensation even in flat space. There is a Fermi surface at $p_F=\mu$.
The scalar frequencies are 
\begin{equation}
\omega_S=\{ \sqrt{p^2+m^2}\pm 2\mu,\ \sqrt{p^2+m^2},\ \sqrt{p^2+m^2},\ p ,\ p \}\ ,
\label{adgh}
\end{equation}
which are always positive definite for $\mu<m/\sqrt{2}$.
The problem is that  BE condensation reappears in an infinitesimal vicinity of $v_x=v_y=v_z=0$.
After turning on $v_z$ and $v_x$, the last frequency in (\ref{adgh}) becomes 
$$
\omega_S= \sqrt{p^2+4 g^2 m^2 v_z^2-2 g m^2 v_x}\ ,
$$
which becomes complex at low momenta in the region $ v_x> 2 g v_z^2$.
Thus the model is not protected from BE condensation.

\end{itemize}

It should be noted that there are many $\mathcal N=1$ supersymmetric models admitting  superconducting phases where the favored classical
vacuum is a SSB vacuum with
$v\neq 0$. But this is a classical effect, it is not BCS superconductivity. In particular, it is not triggered by fermion interactions that become marginal
near a Fermi surface. There are also models containing Fermi surfaces, i.e. regions in parameter space where the fermion
frequencies vanish or take a minimum value at some fermion momentum but, for the reasons explained above, BCS type phase transitions  do not occur generically.

\bigskip

\section{Conclusions}

Understanding how superconducting transitions can take place 
in supersymmetric field theories in full detail is of great interest, in particular, to clarify how supersymmetric theories
react upon the introduction of chemical potential. Other motivations include  providing
a field-theoretical
understanding of the possible mechanisms underlying holographic superconductivity,
and possible applications in real condensed matter systems containing fermion and scalar quasiparticle excitations.
A chemical potential typically leads to Fermi surfaces for fermion fields, and to Bose-Einstein condensation for scalar fields.
In a supersymmetric theory, fermion and scalars are combined with very specific couplings, which have a significant
incidence  in the radiative corrections that determine the one-loop effective potential.
What is the impact of a supersymmetric combination of bosons and fermions 
on phase transitions in thermodynamical systems with chemical potential?
What are the resulting phase diagrams? 

\medskip

To address these questions, in this paper different models have been investigated in detail.  The salient aspects of this investigation
are as follows. The main obstacle to implement BCS superconductivity in a neat way is, as expected,
Bose-Einstein condensation. The model of section   3.1 exhibits the typical problems that arise.
Introducing chemical potential for a baryonic $U(1)_B$ symmetry 
leads to the emergence of Fermi surfaces, but inevitably couples scalar fields
to the chemical potential as well, since scalar fields have the same baryon charge as fermions.
Supersymmetry prevents the scalar  mass scale from getting separated from the fermion mass scale.
Even if supersymmetry is spontaneously broken, the vanishing supertrace relation still implies that there must
be light scalar fields. 
As a result, near the Fermi surface, the  contributions to the thermal potential coming from scalar fields become ill-defined. 
In order to avoid Bose-Einstein   condensation near the Fermi surface one can try to put
 the theory on $S^3$ of radius $R$. This gives an extra mass term $O(1/R)$ to the scalar field 
coming from the coupling to the curvature.  However, because 
the quantized fermion momentum is of order $1/R$, the existence of Fermi surfaces requires chemical potentials that are also corrected by an amount of order $1/R$.
In this model we found that the Fermi surface cannot be separated from the regions of BE condensation.

\medskip

To circumvent these problems, in section 3.2 we  proposed a specific supersymmetric model which realizes BCS superconductivity in flat space. 
It is  based on a K\" ahler potential with quartic terms in the superfields.
\be
K=X^\dagger X +Y^\dagger Y + g^2 (X^\dagger X )^2+ g^2 (Y^\dagger Y )^2\ ,
\ee
and $W=0$. The chemical potential is introduced for  a  $U(1)_R$ symmetry under which  the scalar components
of the $X$ and $Y$ fields have vanishing charge. In this way Bose-Einstein condensation does not occur and the model can be 
studied directly on ${\bf R}^4$. 
This is presumably the simplest supersymmetric model for BCS superconductivity that one can construct, since it contains the minimum number of superfields to have a Dirac fermion, i.e. two chiral superfields,
it can be studied on ${\bf R}^4$ and the superpotential is $W=0$.
We found that the system has a  superconducting phase transition below
some critical temperature, produced by a fermion condensate.

\medskip

The equations determining the temperature dependence of the gap are very similar
to BCS theory, with the main difference represented by the contribution coming from scalar fluctuations. 
 One important effect of this contribution is a drastic reduction  of the dependence on the UV cutoff
 from quadratic to logarithmic. Another effect due to the scalar superpartner is changing the character of the  phase transition from second to first-order.
As explained, the origin of the change is the contribution of low momentum scalar modes at small $\Delta$. 
As the chemical potential is decreased,  $d\Delta/dT$ at $\Delta=0$ becomes large and the phase transition approaches to a second-order phase transition. 

\medskip

In the zero temperature limit, the gap can be computed analytically. The resulting
expression for $\Delta(0)$ is qualitatively similar to
the standard BCS expression. The BCS formula is usually derived by integrating over momenta in a small neighborhood of the Fermi surface, otherwise one does not get a sensible result
--for example, if the Debye energy was  significantly larger than the Fermi energy,
 the formula would be different (for real materials, $|\omega_{Debye}-\omega_{F}|\ll \omega_{F}$).
In the supersymmetric case, due to the cancellation of quadratic divergences,  $\Delta(0)$
can be derived by integrating momenta in the whole range, from $p=0$  up to the cutoff energy, which may  be much larger than the Fermi energy. In short, the supersymmetric expression turns out to be  stable under variation  of parameters.
We have also computed the specific heat. In the superconducting phase, this exhibits the expected exponential suppression due to the gap.

\medskip

An interesting open problem is the construction of supersymmetric BCS models with local $U(1)$ symmetry. In particular,
this would permit the study of magnetic screening and more generally the response of the supersymmetric  system to  external sources.

\section*{Acknowledgements}

We are very grateful to Diego Rodr\'\i guez-G\' omez for collaboration at an early stage of this work, and for  many useful discussions and important insights.
A.B. is supported by a MEC FPU Grant No AP2009-3511.
J.R. acknowledges support by MCYT Research
Grant No.  FPA 2010-20807 and Generalitat de Catalunya under project 2009SGR502.



\renewcommand{\thesection}{\Alph{section}}
\setcounter{section}{0}

\section{Thermal one-loop potential}\label{appA}

In the absence of chemical potential, the one-loop potential at finite temperature is computed 
by using the following formulas (see \cite{RodriguezGomez:2011vg} for a recent discussion and references therein):
\be
V= V_{\rm cl} + V_{\rm CW} + V_{\rm thermal}\ ,
\label{bomo}
\ee
where    $V_{\rm cl}$   represents the classical contribution and
\begin{equation}
\label{VCW}
V_{\rm CW}=\frac{1}{64\pi^2}\sum_i (-1)^F M^4_i\log \frac{M^2_i}{\Lambda^2}\ ,
\end{equation}

\begin{equation}
\label{DeltaV}
V_{\rm thermal}=\frac{1}{\beta}\,\sum_i (-1)^F\int \frac{d^3\vec{p}}{(2\pi)^3}\, \log\Big(1- (-1)^F\, e^{-\beta\,E_i} \Big)\ ,
\end{equation}
where $E^2_i=\sqrt{\vec{p}\,^2+M_i^2}$ and the sum over $i$ accounts for  scalar  ($F=0$) and  fermion ($F=1$) degrees of freedom,
and $M_i$ are the mass eigenvalues of the scalar and fermion mass matrices of the quadratic fluctuations, and
 $\Lambda $ represents a UV cutoff.
By expanding the logarithm in (\ref{DeltaV}) the integral over $p$ can be carried out explicitly (see e.g.
\cite{RodriguezGomez:2011vg}). At sufficiently small temperatures, the contribution from massless modes dominate over the massive mode contributions, which become exponentially suppressed.
This has the important implication that the SSB  vacuum, which contains a massless mode --the Goldstone boson-- will necessarily dominate
the thermodynamics at low temperatures.

When chemical potential is added into the system, the one-loop determinant is given in terms of eigenvalues of the frequencies.
 The complete one-loop potential is given by (\ref{bomo}) with
\begin{gather}
V_{\rm CW} =\frac{1}{2}\sum_i\int \frac{d^3\vec{p}}{(2\pi)^3}\omega_{Si}-\frac{1}{2}\sum_i\int \frac{d^3\vec{p}}{(2\pi)^3}\omega_{Fi}\
,\\
V_{\rm thermal}=\frac{1}{\beta}\,\sum_i \int \frac{d^3\vec{p}}{(2\pi)^3}\, \log\Big(1-\, e^{-\beta\,\omega_{Si}} \Big)-\frac{1}{\beta}\,\sum_i \int \frac{d^3\vec{p}}{(2\pi)^3}\, \log\Big(1+\, e^{-\beta\,\omega_{Fi}} \Big)\
, 
\end{gather}
where $\omega_{Si}$ and $\omega_{Fi}$ are the scalar and fermion frequencies.
We recall the origin of these expressions.
In the Euclidean theory, time is periodic and one needs to impose periodic and antiperiodic boundary conditions for scalars and fermions, respectively. This leads to 
quantized frequencies,
\begin{equation}
\omega_S=\frac{2\,\pi\,n}{\beta}\ ,\qquad \omega_F=\frac{\pi\,(2\,n+1)}{\beta}\ ,
\end{equation}
with integer $n$. In particular, for scalars,  one has to compute
\begin{equation}
\sum_i \frac{1}{2\beta}\sum_n\,\log (\omega_S^2+\omega^2_{Si})\ .
\end{equation}
The sum over $n$ can be easily performed by first differentiating with respect to $\omega_{Si}^ 2$. One finds
\begin{equation}
\frac{1}{2\beta}\sum_n\,\log (\Big(\frac{2\,\pi\,n}{\beta}\Big)^2+\omega^2_{Si})=\frac{1}{\beta}\,\log\sinh\frac{\beta\, \omega_{Si}}{2}=\frac{\omega_{Si}}{2}+\frac{1}{\beta}\log\left(1-e^{-\beta\omega_{Si}}\right)\ ,
\label{sumas}
\end{equation}
plus a constant which we  discard. 

A similar calculation for the fermions gives
$\frac{\omega_{Fi}}{2}+\frac{1}{\beta}\log\left(1+e^{-\beta\omega_{Fi}}\right)\ 
$.

Consider now the construction of the one-loop potential for the theory on $S^1\times S^3$.
As explained, once we  put the theory on  $S^3$,  the integral over momentum is replaced by a discrete sum over Kaluza-Klein modes of $S^ 3$.
The quantized momenta are as follows 
\bea
&&{\rm Scalars:}\qquad  p^2\ \longrightarrow \  l (l+2)R^{-2} \ ,
\nn\\
&&{\rm Fermions:}\ \ \  p^2\ \longrightarrow \ (l+1/2)^2 R^{-2}\ ,
\eea
with $l=0,1,2,...$. Thus
we have the  prescription
\be
\int\frac{d^3\vec{p}}{(2\pi)^3}\to\frac{1}{{\rm Vol}\,S^3} \sum_{l=0}^ \infty d_l\ , 
\ee
where $d_l$ represents the scalar degeneracy  $d_l^S= (l+1)^ 2$, or fermion degeneracy ,  $d_l^F= l(l+1)$. 
Thus, the  one-loop corrected effective potential is
$\Omega=V_{\rm cl}+V_{\rm CW}+V_{\rm thermal}$ where 
 $V_{\rm cl}$ is the classical potential and 
\bea
V_{\rm CW} &=&\frac{1}{2}\frac{1}{{\rm Vol}\,S^3}\sum_i\, \sum_{l=0}^ \infty \,( d_l^S\omega_{Si}-  d_l^F\omega_{Fi})\ ,
\nn\\
V_{\rm thermal}&=& 
\frac{1}{\beta}\frac{1}{{\rm Vol}\,S^3}\,\sum_i\,
 \sum_{l=0}^ \infty \left( d_l^S
 \,\log (1-e^{-\beta\,\omega_{Si}}) -  d_l^F \log(1+e^{-\beta\,\omega_{Fi}}) \right)\ .
\eea
The sum over $l$ in the thermal part $V_{\rm thermal}$  is, as usual, convergent.
However, the sum over $l$ in $V_{\rm CW}$ diverges.
 Following \cite{Hollowood:2008gp},
 we regularize it by putting a momentum cutoff in the maximum allowed  energy:
$$
\sum_{l=0}^\infty h(l)\ \longrightarrow \ \sum_{l=0}^\infty h(l)\tilde\theta (\omega_l/\Lambda)\ ,
$$
where $\tilde\theta(x)$ is 1 for $x<1$ and 0 for $x>1$, and $\omega_l$ is the energy of the (scalar or fermion)  mode. Note that the cutoff
is on the energy, not on $l$, so the $\tilde \theta $ function is slightly different for scalars and fermions (but such that that the energy of the last
mode to be included in the sum is the same for both scalars and fermions). Then we replace the sum over $l$ by integrals by using the Abel-Plana formula
\be
\sum_{l=0}^\infty F(l) = \int_0^\infty dx\  F(x)+{1\over 2}F(0)-2\int_0^\infty dx \ \frac{{\rm Im}F(ix)}{e^{2\pi x}-1}\ .
\ee
The divergence in the sum over $l$ is reflected in the first term. The integral will give rise to a $\Lambda^4R^ 3$ piece which cancels
between scalars and fermions, because this term  is simply multiplied by the number of degrees of freedom which is the same for scalars and fermions.
Then there is a $\Lambda^ 2 R$ term. In flat space, this term cancels out  due to the vanishing  supertrace formula. On $S^3$ (as noticed in \cite{Hollowood:2006xb}) it does not cancel out.
However, its coefficient is a constant independent of $\mu$ and independent of $v$. Therefore it will not affect the vacuum dynamics (which depends on the difference
of free energies). Finally, there is the expected  term proportional to $\log\Lambda $. In a renormalizable theory, this can be canceled 
by the addition of a suitable counterterm. In the present case, we have an effective field theory model which is valid up to the energy scale  $\Lambda$ set by the dimensionful parameter $g^{-1}$ (analogous to Fermi energy
in the Fermi theory of weak interactions).



\section{Kinetic term matrices}\label{appB}

We write   the quadratic terms in the fluctuation Lagrangian as $\Phi^\dagger O_S\Phi$ for the scalar part, and
 as $\Psi^\dagger O_F\Psi$ for the fermion part, where
\begin{gather}
\Phi^\dagger=(\phi_x^*(p),\phi_y^*(p),\phi_x(-p),\phi_y(-p))\,,\\
\Psi^\dagger=(\psi_{x1}^\dagger(p),\psi_{x2}^\dagger(p),\psi_{y1}^\dagger(p),\psi_{y2}^\dagger(p),\psi_{x1}(-p),\psi_{x2}(-p),\psi_{y1}(-p),\psi_{y2}(-p))\,.
\end{gather}

Consider first the Lagrangians \eqref{LBUBflat}, 
\eqref{LFUBflat}, describing the model where 
 the chemical potential is coupled to a $U(1)_B$ current.
When $v_x=0,\, v_y=0$, the kinetic term matrices, $O_S$ and $O_F$,  take the following simple
form:
\begin{multline}
O_S=\frac{1}{2}\,{\rm diag}\big[\left((\omega+i\mu) ^2+p^2+m^2+4 g^4 \Delta _x^2\right),\left((\omega-i\mu) ^2+p^2+m^2+4 g^4 \Delta _x^2\right),\\
\left((\omega-i\mu) ^2+p^2+m^2+4 g^4 \Delta _x^2\right),\left((\omega+i\mu) ^2+p^2+m^2+4 g^4 \Delta _x^2\right)\big]
\end{multline}
\begin{equation}
O_F=\left(
\begin{array}{cc}
A_+&B\\
-B&A_-
\end{array}
\right)
\end{equation}
\begin{equation}
 A_\pm=\left(
\begin{array}{cccc}
\frac{1}{2} (i\omega-p\mp\mu  ) & 0 & 0 & 0\\
0 & \frac{1}{2} (i\omega+p\mp\mu  ) & 0 & 0\\
0 & 0 & \frac{1}{2} (i\omega-p\pm\mu  ) & 0\\
0 & 0 & 0 & \frac{1}{2} (i\omega+p\pm\mu  ) \\
\end{array}
\right)
\end{equation}
\begin{equation}
B=\left(
\begin{array}{cccc}
 0 & -g^2 \Delta _x & 0 & \frac{m}{2} \\
 g^2 \Delta _x & 0 & -\frac{m}{2} & 0 \\
 0 & \frac{m}{2} & 0 & -g^2 \Delta _y \\
 -\frac{m}{2} & 0 & g^2 \Delta _y & 0 \\
\end{array}
\right)
\end{equation}
We omit the (long)  general expressions with $v_x, v_y\neq 0$ as these are not used in the discussion.

{} For the  Lagrangian \eqref{LBUR} and \eqref{LFUR}, corresponding to a  chemical potential coupled to a $U(1)_R$ symmetry, we get the
following expressions
\begin{equation}
O_S=\left(
\begin{array}{cc}
 \frac{1}{2}\left(\omega ^2+p^2\right)
+\frac{2g^2\Delta^2}{1+4g v^2} & 0 \\
 0 & \frac{1}{2}\left(\omega ^2+p^2\right)+\frac{2g^2\Delta^2}{1+4g v^2} \\
\end{array}
\right) \, ,
\end{equation}

\begin{equation}
O_F=\left(
\begin{array}{cccc}
 \frac{1}{2}(i \omega -p)+\frac{\mu }{2} & 0 & 0 & -\frac{g \Delta }{1+4g v^{2}} \\
 0 & \frac{1}{2}(i \omega +p)+\frac{\mu }{2} & \frac{g \Delta }{1+4g v^{2}} & 0 \\
 0 & \frac{g \Delta }{1+4gv^2} & \frac{1}{2}(i \omega -p)-\frac{\mu }{2} & 0 \\
 -\frac{g \Delta }{1+4gv^2} & 0 & 0 & \frac{1}{2}(i \omega +p)-\frac{\mu }{2} \\
\end{array}
\right)\,.
\end{equation}
Here $\{ \Delta , v\} $ stands for $\{ \Delta_x , v_x\} $ or  $\{ \Delta_y , v_y\} $ as applied to the kinetic matrix involving  the $X$ or the $Y$ superfield components, 
 as the two chiral superfields $X$ and $Y$ are decoupled.


\end{document}